\documentclass[twocolumn,prd,superscriptaddress,showpacs,floatfix,%
preprintnumbers,nofootinbib]{revtex4}

\usepackage{epsfig}
\usepackage{bm}

\begin{document}

\title[Non-Universal Spectra of Ultra-High Energy Cosmic Rays in a Structured Universe]{Non-Universal Spectra of Ultra-High Energy Cosmic Ray Primaries and Secondaries in a Structured Universe}

\author{G{\"u}nter Sigl}
\affiliation{APC~\footnote{UMR 7164 (CNRS, Universit\'e Paris 7,
CEA, Observatoire de Paris)} (AstroParticules et Cosmologie),
10, rue Alice Domon et L\'eonie Duquet, 75205 Paris Cedex 13, France\\
and Institut d'Astrophysique de Paris, UMR 7095 CNRS - Universite Pierre \&
Marie Curie, 98 bis boulevard Arago, F-75014 Paris, France}

\begin{abstract}
Analytical calculations of extra-galactic cosmic ray spectra above 
$\sim10^{17}\,$eV are
often performed assuming continuous source distributions, giving rise
to spectra that depend little on the propagation mode, be it rectilinear
or diffusive. We perform trajectory simulations for proton primaries
in the probably more realistic case of discrete sources with a density of
$\sim10^{-5}\,{\rm Mpc}^{-3}$. We find two considerable non-universal
effects that depend on source distributions and magnetic fields:
First, the primary extra-galactic cosmic ray flux can become strongly
suppressed below a few $10^{18}\,$eV due to partial confinement in magnetic
fields surrounding sources. Second, the
secondary photon to primary cosmic ray flux ratio between
$\simeq3\times10^{18}\,$eV and $\simeq10^{20}\,$eV
decreases with decreasing source density and increasing magnetization.
As a consequence, in acceleration scenarios for the origin of
highest energy cosmic rays the fraction of secondary photons may be
difficult to detect even
for experiments such as Pierre Auger. The cosmogenic neutrino flux
does not significantly depend on source density and magnetization.

\end{abstract}



\maketitle

\section{Introduction}
A major unresolved aspect of ultra-high energy cosmic ray (UHECR) physics~\cite{Cronin:2004ye} is their composition and above which energy the flux is dominated by extragalactic sources.
Above $\simeq10^{17}\,$eV the chemical composition is basically unknown~\cite{Watson:2004rg}. Around $10^{18}\,$eV the situation is particularly inconclusive as HiRes~\cite{Abbasi:2004nz} and HiRes-MIA~\cite{Abu-Zayyad:2000ay} data suggest a light (proton dominated) composition, whereas other experiments indicate a heavy composition~\cite{Hoerandel:2004gv}.

As a consequence, there are currently two different scenarios: The "standard" one,
where a transition from a steeper, galactic heavy component to a flatter, extragalactic component dominated by protons takes place at the ankle at $\simeq5\times10^{18}\,$eV, see, e.g., Ref.~\cite{Wibig:2004ye},
and a more recent one suggesting that this transition actually takes place at lower energies, namely around the "second knee" at $\simeq4\times10^{17}\,$eV.

This second scenario in which extragalactic protons dominate down to the second knee has the following consequences: First, since the observed spectrum above the second knee is quite steep, $\propto E^{-3.3}$, the extragalactic proton flux has to cut off below $\simeq4\times10^{17}\,$eV. This can be explained as a magnetic horizon effect: Protons from cosmological distances cannot reach the observer any more within a Hubble time due to diffusion in large scale magnetic 
fields~\cite{Isola:2001ng,Stanev:2001rr,Stanev:2003hg,Deligny:2003rp,Lemoine:2004uw,Aloisio:2004fz}.
Second, the ankle would have to be interpreted as due to pair production of the extragalactic protons~\cite{Berezinsky:2002nc,Berezinsky:2005cq}. In particular, it has been pointed out recently~\cite{Berezinsky:2005cq,Allard:2005ha}, that this model cannot afford injection of a significant heavy component above the ankle whose secondary photo-disintegration products in the form of intermediate mass nuclei would produce a bump around the ankle. Injection of a mixed composition would also require a harder injection spectrum
$\propto E^{-\alpha}$ with $\alpha\simeq2.2$, as opposed to the extragalactic ankle scenario with pure protons~\cite{Berezinsky:2005cq} which requires an injection spectrum $\propto E^{-2.6}$.

In the proton dominated low-cross over scenario the resulting spectra are
surprisingly insensitive to details such as extra-galactic magnetic
fields (EGMF) and actual source distribution, as long as
the distances between UHECR sources are much smaller than the
energy loss and diffusion lengths~\cite{Aloisio:2004jd,Aloisio:2006wv}.
The spectra are then equal
to the spectra resulting from a homogeneous source distribution
which is therefore called ``universal''. Continuously distributed sources
fulfill this condition.

Actual UHECR sources are, however, likely discrete with a relatively
small density of order $\sim10^{-5}\,{\rm Mpc}^{-3}$~\cite{Kachelriess:2004pc}.
These values are motivated by the density of candidate sources such
as active galaxies and also by hints of UHECR clusters.
In addition, there likely is a strongly structured EGMF. In the present
paper we study how a strongly structured Universe can cause deviations
from the universal spectra in the case where extragalactic protons
dominate the observed flux down to $\sim10^{18}\,$eV.

We also include in this study secondary neutrinos and
$\gamma-$rays, produced by interactions of the primary nucleons with
the cosmic microwave background (CMB) and the infrared (IR) background.
The relevant reactions which also govern energy loss of the primary
nucleons are pion production above
the ``GZK threshold''~\cite{gzk} and, for $\gamma-$rays, by pair production
of protons. There is a strong motivation to study predictions for the
fraction of $\gamma-$rays in the UHECR flux from recent experimental
upper limits~\cite{Shinozaki:2002ve,Risse:2005jr,Rubtsov:2006tt,Abraham:2006ar,Glushkov:2007ss} 
that will further improve in the near future, especially from the Pierre Auger project~\cite{Risse:2007sd}. Even secondary neutrino fluxes
may be detectable in the not too far future~\cite{McDonald:2003xn}.

In section 2 we describe our simulations, in section 3 we discuss
results for various degrees of structure in the source distribution
and magnetic field scenarios. In section 4 we discuss uncertainties
in the $\gamma-$ray to charged primary flux ratio predicted in
bottom-up scenarios and we conclude in section 4.
We use natural units, $\hbar=c=k=1$, throughout the paper.

\section{Numerical Simulations}
For the trajectory simulations we use the public code package
CRPropa~\cite{crpropa,Armengaud:2006fx}. It follows nucleon trajectories
in 3 dimensions in arbitrarily structured environments and takes into
account secondary neutrinos and electromagnetic (EM)
cascades produced by nucleon interactions with the CMB and IR
backgrounds. Note that
trajectory simulations allow to treat both the rectilinear and
diffusive regime without approximation and are thus more general
than using the diffusion approximation as adopted in Ref.~\cite{Aloisio:2004jd,Aloisio:2006wv} which in addition was
restricted to a homogeneous EGMF. We will also compare the general
case to the isotropic situation in the absence of UHECR deflection
and source structure which can be simulated with the
one-dimensional version of CRPropa in which primary and secondary
particles just propagate along straight lines. This case has been
considered, e.g., in Ref.~\cite{Gelmini:2005wu,Gelmini:2007sf}.

We inject protons with a spectrum $E^{-\gamma}$ up to $10^{21}\,$eV and
adjust the injection index $\gamma$ to the data. We also consider scenarios
where the injection power per comoving volume evolves as a power law
$\propto(1+z)^m$, where $m$ is often called the ``bright phase index''.
We integrate up to redshift $z=3$.

For nucleons we take into account photo-pion production, and 
pair production on the CMB and IR background, as well as
redshift, and deflection in the cases where we consider an EGMF.

The $\gamma-$ray interaction length above $\sim10^{19}\,$eV strongly
depends on the density of the universal radio background 
(URB)~\cite{Bhattacharjee:1998qc}. We
use the minimal estimate for the URB~\cite{Protheroe:1996si}.
For the IR background we use the model of Primack et 
al.~\cite{Primack:2005rf}. The IR background is mostly relevant for
pion production for relatively steep proton injection spectra, and
for photon attenuation in the TeV regime. Since 
we are mostly interested in energies above $\sim10^{17}\,$eV, for
our purposes the IR background is less relevant for EM cascade
propagation. Between
$\sim10^{17}\,$eV and $\sim10^{21}\,$eV, the $\gamma-$ray attenuation
length is smaller than 10 Mpc, and thus redshift effects are negligible
at these energies.
Furthermore, since we are interested in diffuse fluxes, we can treat
the EM cascades as 1-dimensional.

Neutrinos propagate along straight lines and, once produced, are only subject
to redshift.

The concordance cosmology is used for which,
assuming a flat Universe, the Hubble rate $H(z)$ at redshift $z$
in the matter dominated regime, $z\lesssim10^3$, is given by
$H(z)= H_0\left[\Omega_{\rm m}(1+z)^3+\Omega_{\Lambda}\right]^{1/2}\,$.
We use the standard values being $\Omega_{\rm m}=0.3$,
$\Omega_{\Lambda}=0.7$, and $H_0=h_0\,100~{\rm km}~{\rm s}^{-1}~{\rm
Mpc}^{-1}$ with $h_0=0.72$.

\section{Results}

\begin{figure}[ht]
\includegraphics[width=0.5\textwidth,clip=true]{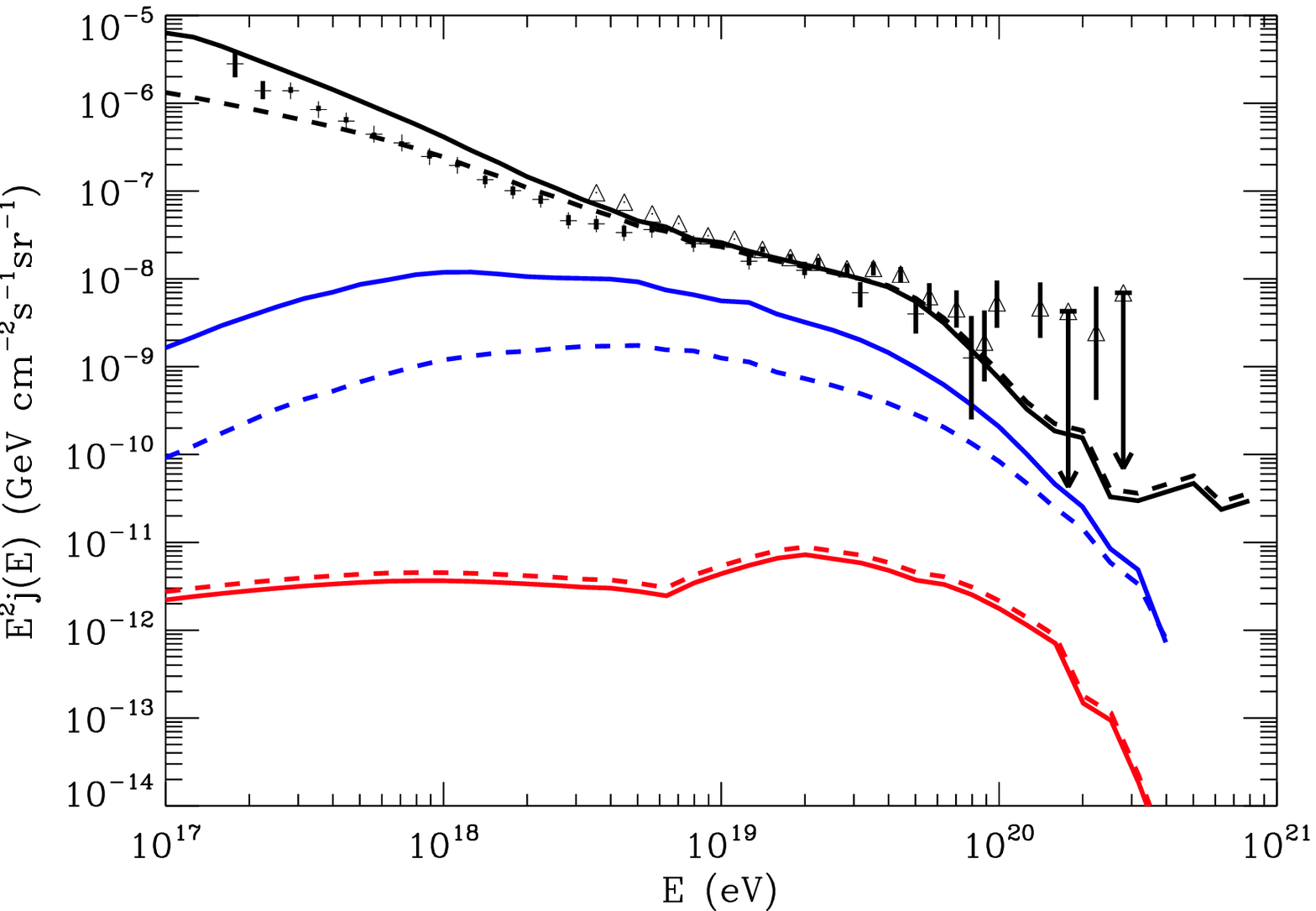}
\includegraphics[width=0.5\textwidth,clip=true]{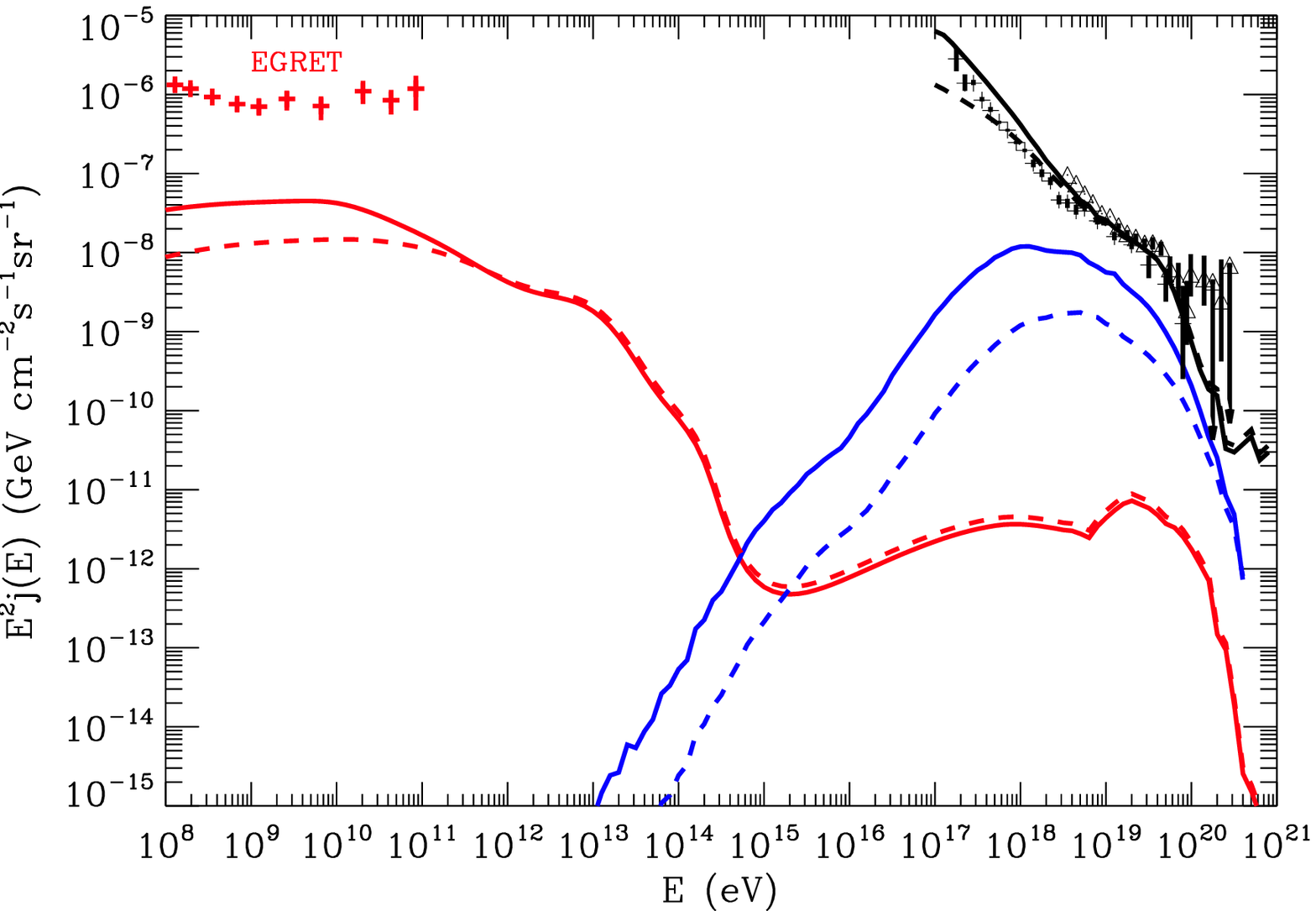}
\caption{Upper panel: Fluxes of primary nucleons (black), secondary
photons (red) and neutrinos per flavor (blue) for a homogeneous source distribution injecting an $E^{-2.6}$ proton spectrum up to $10^{21}\,$eV
and up to redshift $z=3$. The dashed lines are in absence of redshift
evolution, the solid lines are for a bright phase index $m=3$. No magnetic
fields are taken into account and fluxes have been computed with the one
dimensional version of CRPropa. AGASA data~\cite{Shinozaki:2006kk} are shown as
triangels, HiRes data~\cite{Abbasi:2002ta} as crosses. Lower panel: Same
over a wider energy range. Also shown is the diffuse $\gamma-$ray flux
that has been observed by EGRET~\cite{Strong:2003ex}.}
\label{fig1}
\end{figure}

\begin{figure}[ht]
\includegraphics[width=0.5\textwidth,clip=true]{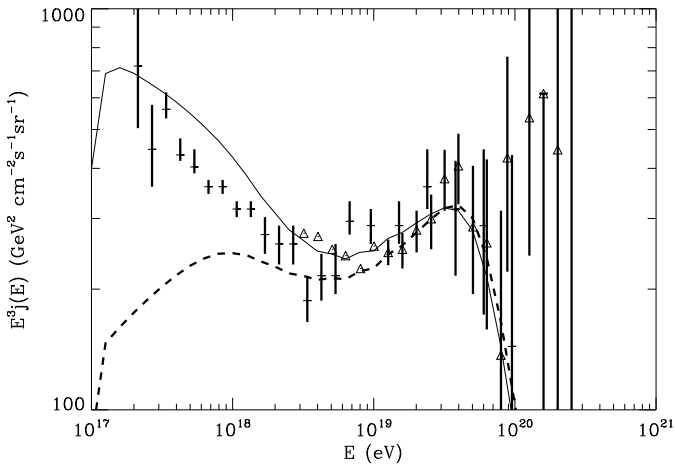}
\caption{Same as Fig.~\ref{fig1}, upper panel, but with the energies
measured by HiRes and AGASA multiplied by a factor of 1.2 and
0.9, respectively, and showing the differential nucleon flux multiplied
by $E^3$.}
\label{fig2}
\end{figure}

In Fig.~\ref{fig1} we show nucleon, $\gamma-$ray and neutrino fluxes
for a Universe filled with homogeneously distributed sources, with
and without source evolution. As pointed out in 
Ref.~\cite{Berezinsky:2002nc,Aloisio:2006wv}, the HiRes and AGASA
spectra can be made to give a consistent ankle structure, coincident with the
theoretically predicted ankle position at $\simeq5\times10^{18}\,$eV,
by multiplying the HiRes energies with a factor 1.2 and the AGASA
energies with a factor 0.9. 
This is best seen by multiplying the differential spectrum with $E^3$,
as shown in Fig.~\ref{fig2}. The best fit to the observed flux
down to $\simeq2\times10^{17}\,$eV
then gives an injection spectral index of $2.6\lesssim\gamma\lesssim2.7$,
where the lower value corresponds to $m\simeq3$ and the upper value to
$m\simeq0$.

As the lower panel of Fig.~\ref{fig1} of shows, the $\gamma-$ray flux at GeV 
energies can reach about ten percent of the diffuse background observed by 
EGRET~\cite{Strong:2003ex} in this scenario, particularly for significant source 
evolution. The energy in GeV
$\gamma-$rays is comparable to the energy in primary cosmic rays at 
$\simeq2\times10^{18}\,$eV, where the proton energy attenuation length
due to pair production becomes comparable to the Hubble 
radius~\cite{Bhattacharjee:1998qc}. A diffuse $\gamma-$ray flux at this
level should be testable in the near future~\cite{semikoz-sigl}.

\begin{figure}[ht]
\includegraphics[width=0.5\textwidth,clip=true]{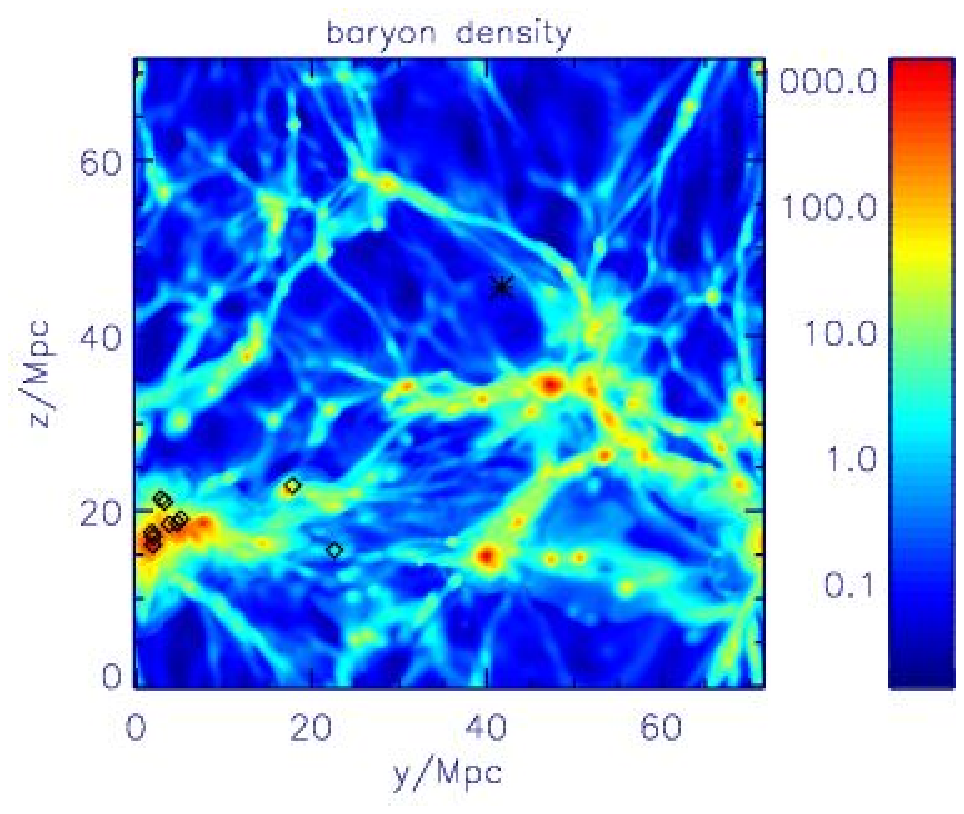}
\includegraphics[width=0.5\textwidth,clip=true]{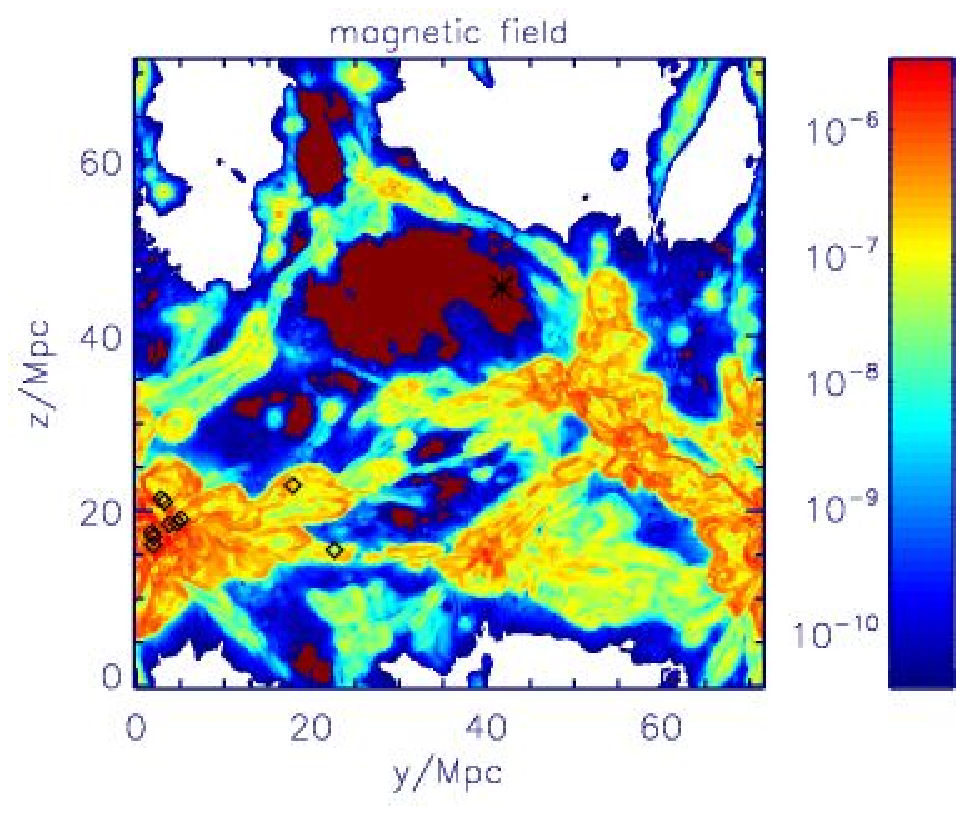}
\caption{Cross-section through the LSS simulation box from
Refs.~\cite{ryu,miniati} used in this study. The observer position
is marked as a black cross near the center. The approximate
positions of the ten sources in the discrete source realization
investigated below are shown as black diamonds. Note that the
simulation box including EGMF and sources is periodically repeated
in all three directions. Upper panel: Baryon density.
Lower panel: magnetic field strength for the EGMF considered.
The field polarizations are not shown and are in general
a combination of smooth and stochastic components.}
\label{fig3}
\end{figure}

As in our earlier work in Refs.~\cite{sigl-miniati},
for the scenarios with structured source distributions and/or
magnetic fields we will in the following use the unconstrained
large scale structure (LSS) simulations based on Refs.~\cite{ryu,miniati}.
A cross-section through the baryon density and magnetic field strength
of the LSS simulation box is shown in Fig.~\ref{fig3}. The EGMF in these
simulations have been evolved passively and normalized to $\sim\mu$Gauss
in the centers of galaxy clusters.
These EGMF are highly structured in that they reach a few microGauss
in the most prominent structures such as galaxy clusters, but is
$\lesssim10^{-11}\,$G in the voids. We note that the EGMF in this
LSS simulation are relatively extended compared to other
simulations~\cite{dolag} and current observations do not allow to
distinguish between such different EGMF scenarios. The EGMF obtained in Ref.~\cite{dolag}
is closer to the case of negligible EGMF that we also study here.

Since the LSS simulation covers a volume of only $\simeq(75\,{\rm Mpc})^3$,
the structures and fields in the simulation box are periodically repeated
in all three directions. In this way, no evolution of the structure
and fields with redshift is taken into account. This is sufficient
within the uncertainties in such scenarios since at distances larger
than the GZK distance, spectra and angular distributions are insensitive
to the detailed structure. The observer is chosen in a low magnetic field
region of the simulation box, resembling Earths actual environment, as
can be seen in Fig.~\ref{fig3}.

\begin{figure}[ht]
\includegraphics[width=0.5\textwidth,clip=true]{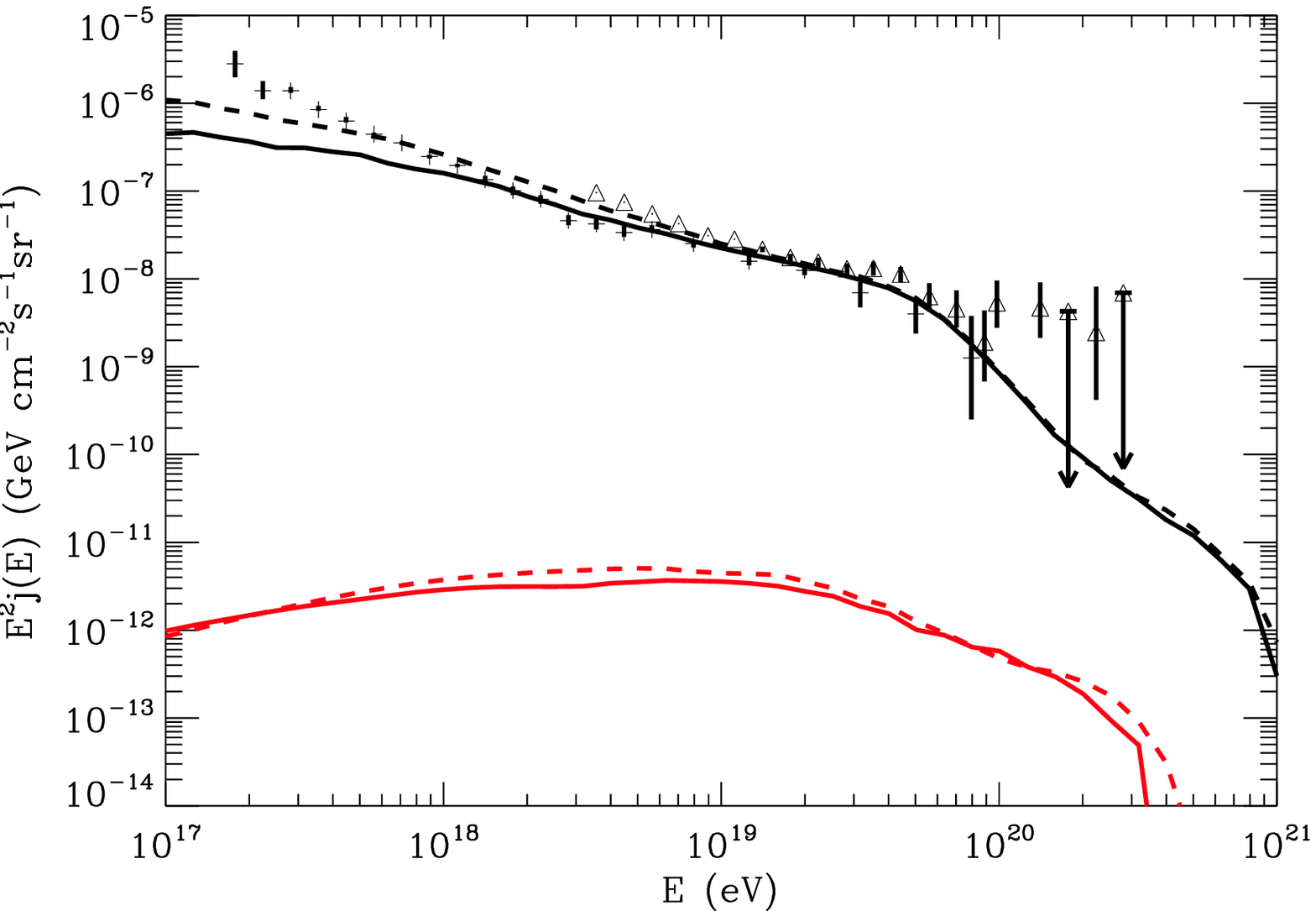}
\includegraphics[width=0.5\textwidth,clip=true]{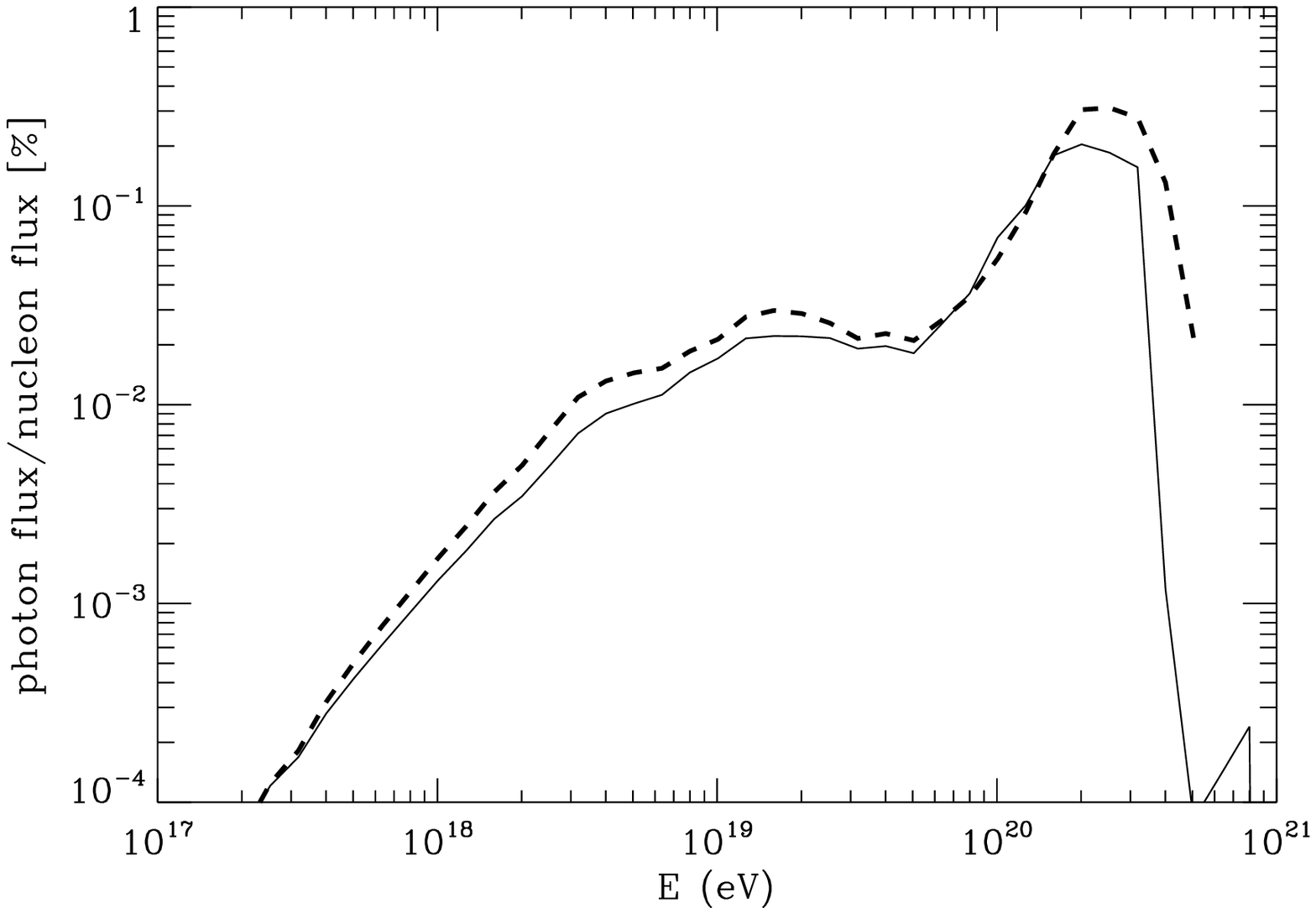}
\caption{Upper panel: Fluxes of primary nucleons (black) and secondary
photons (red) for continuous sources
following the baryon density of the LSS simulation in
Refs.~\cite{ryu,miniati} and injecting an $E^{-2.6}$ proton spectrum
up to $10^{21}\,$eV and redshift $z=3$ with no source evolution, $m=0$.
Solid line assume the EGMF from these LSS simulations, whereas
dashed lines are for negligible EGMF. Data as in Fig.~\ref{fig1}.
Lower panel: Photon to cosmic ray flux ratio for this case.}
\label{fig4}
\end{figure}

Fig.~\ref{fig4} shows results for structured, but continuous sources
whose density is assumed proportional to the baryon density in the
LSS simulation of which a cross-section is shown in Fig.~\ref{fig3}.
This time, the case with negligible EGMF is compared
with the EGMF obtained from the LSS simulations, as in Refs.~\cite{sigl-miniati}.
The spectra are hardly different from the homogeneous
case shown in Fig.~\ref{fig1}. This scenario therefore corresponds to
the universal spectrum discussed in Ref.~\cite{Aloisio:2004jd}. The
proton flux is only slightly suppressed by the EGMF at a few $10^{18}\,$eV.
The neutral fraction, shown in the lower panel of Fig.~\ref{fig4} is
also quite insensitive to both the continuous source distribution and the EGMF. It is at the lower end of the range obtained
in Ref.~\cite{Gelmini:2005wu,Gelmini:2007sf}.

In the following we define the photon to cosmic ray flux ratio as
relative to the total observed cosmic ray flux below $3\times10^{19}\,$eV
and relative to the simulated nucleon flux above $3\times10^{19}\,$eV.
This photon to cosmic ray flux ratio, also shown in Fig.~\ref{fig4},
in fact broadly follows the
attenuation length of the EM cascade~\cite{Bhattacharjee:1998qc}.
This is to be expected as long as the
primary flux does not exhibit a strong break between the energy $E$
at which the secondary to primary ratio is considered and the energies
$E^\prime > E$ of the primaries mostly responsible for the secondary
flux at energy $E$. Below we will see that below $\simeq3\times10^{18}\,$eV
and above $\simeq10^{20}\,$eV, the photon to nucleon ratio indeed
depends little on source distributions and magnetization and mostly
follows the behavior of the EM cascade attenuation length. Since in the
combined CMB and URB this length scale increases roughly linearly
with energy between $\simeq10^{20}\,$eV and 
$\simeq10^{21}\,$eV~\cite{Bhattacharjee:1998qc},
this leads to a rise in the photon to nucleon ratio
up to the order of 1 percent in this energy range.
Unfortunately, due to the sharp drop
of the primary cosmic ray flux expected at these energies, a photon
to nucleon ratio of order percent may not be easier to detect than an
order $10^{-4}$ ratio at lower energies.

\begin{figure}[ht]
\includegraphics[width=0.5\textwidth,clip=true]{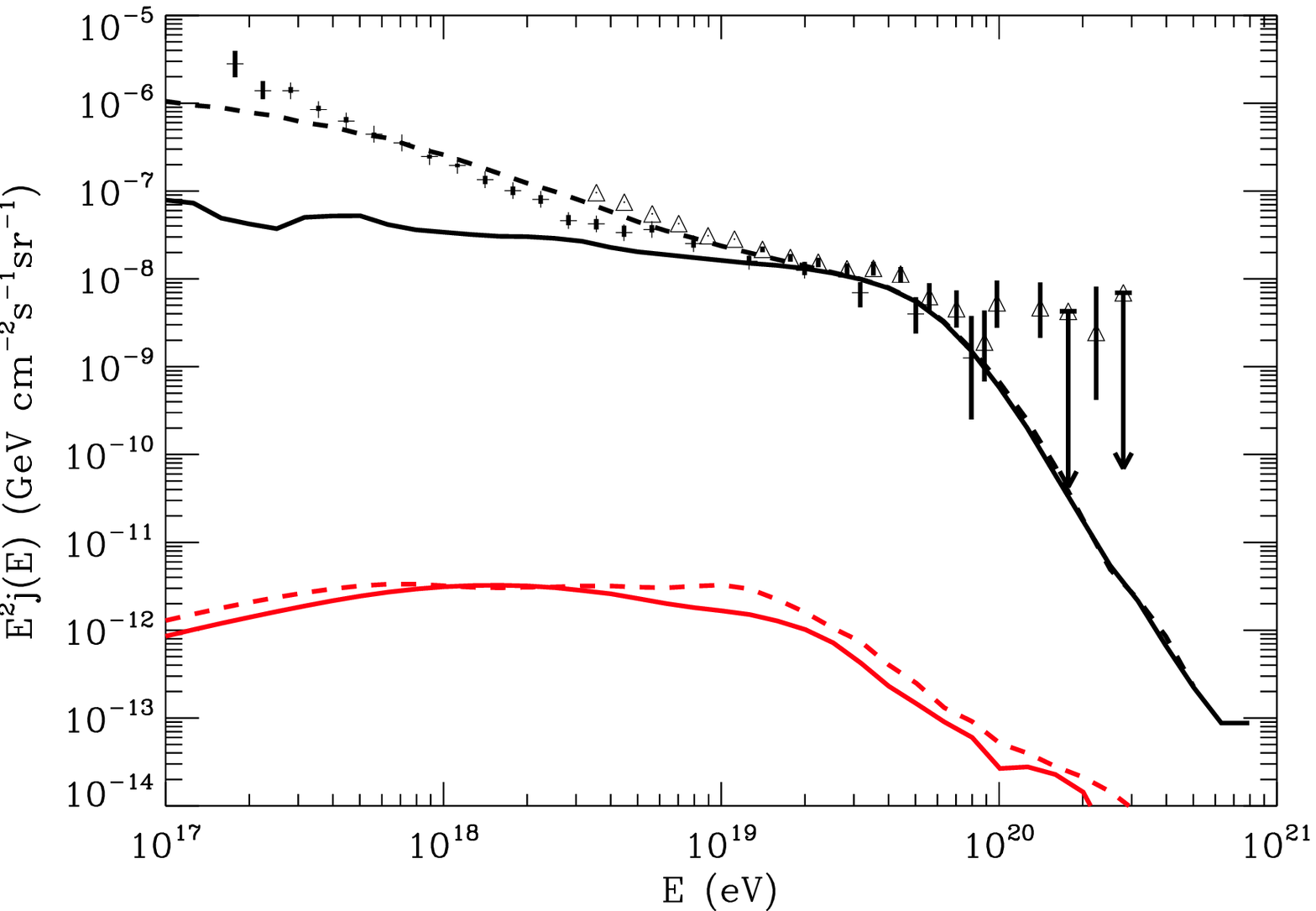}
\includegraphics[width=0.5\textwidth,clip=true]{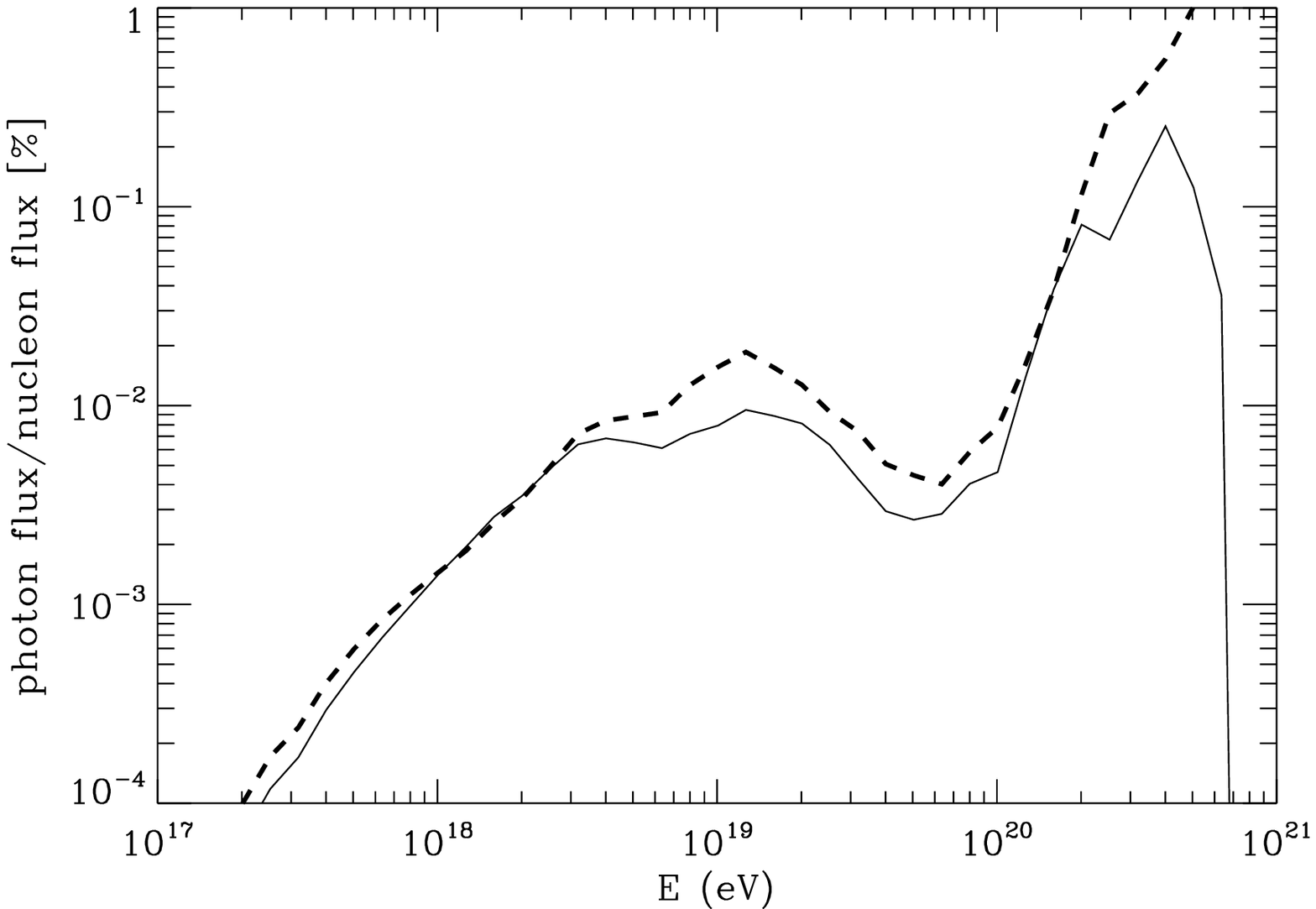}
\caption{Same as Fig.~\ref{fig4} for the discrete source realization
with 10 sources in the LSS simulation box, see Fig.~\ref{fig3},
injecting an $E^{-2.6}$ proton spectrum up to $10^{21}\,$eV and
redshift $z=3$ with no source evolution, $m=0$. This corresponds to a
source density of $\simeq3\times10^{-5}\,{\rm Mpc}^{-3}$. Again,
solid line assume the EGMF from the LSS simulations, whereas
dashed lines are for negligible EGMF, and no source evolution is
assumed.}
\label{fig5}
\end{figure}

We now consider a specific realization with discrete sources of
density $n_s\sim3\times10^{-5}\,{\rm Mpc}^{-3}$. As Fig.~\ref{fig3}
shows, the sources in this realization are concentrated around the
most prominent galaxy cluster. The
spectra resulting in this scenario are shown in Fig.~\ref{fig5},
upper panel. Both above the GZK energy around $10^{20}\,$eV
and below $\simeq3\times10^{18}\,$EeV the primary spectra
are not universal any longer. This is understandable
because the average source distance $\sim30\,$Mpc is now comparable
to the GZK distance. Further, a cosmic ray of energy $E$ in a region
permeated by magnetic fields $B$ has a Larmor radius $r_{\rm L}\sim1\,(E/{\rm EeV})(B/\mu{\rm G})^{-1}\,$kpc and during the lifetime of the Universe,
$t_u\sim10\,$Gyr, diffuses a distance
\begin{equation}
l_d(E)\sim\left(r_{\rm L}t_u\right)^{1/2}\sim2\,\left(\frac{E}{\rm EeV}\right)^{1/2}\left(\frac{B}{\mu{\rm G}}\right)^{-1/2}\,{\rm Mpc}\,,
\label{l_d}
\end{equation}
as long as this length scale is smaller than the spatial extent of
the field. Fig.~\ref{fig3} shows that the structured EGMF reaches values of 
$B\sim0.1\,\mu$G over several Mpc around the sources in the LSS simulation
used here. Below $\sim3\times10^{18}\,$eV, the diffusion length then 
also becomes smaller than the average source distance
\begin{equation}
  d_s\simeq n_s^{-1/3}\simeq46\,
  \left(\frac{n_s}{{\rm Mpc}^{-3}}\right)^{-1/3}\,{\rm Mpc}\,.
  \label{d_s}
\end{equation}
This leads to a suppression of the nucleon flux due to a partial 
confinement of UHECRs which was discussed qualitatively in 
Ref.~\cite{Lemoine:2004uw}. Note
that this is not quite the same as the magnetic horizon effect discussed in 
Refs~\cite{Isola:2001ng,Lemoine:2004uw,Aloisio:2004fz,Berezinsky:2007kz} where
typically uniform EGMFs were assumed which is less realistic in a structured 
universe.

Above the threshold for pair production by photons with the CMB at 
$\simeq10^{15}\,$eV
and up to $\simeq10^{20}\,$eV, the photon attenuation length is smaller
than 10 Mpc~\cite{Bhattacharjee:1998qc}. Photons in this energy range
thus have to be produced within $\lesssim10\,$Mpc from the observer.
Above $\simeq3\times10^{18}\,$eV $\gamma-$rays are mostly produced by nucleons
above the GZK threshold whose attenuation length is smaller than
the typical source distance Eq.~(\ref{d_s}). This explains why the
photon fraction of the UHECR flux between $\simeq3\times10^{18}\,$eV
and $\simeq10^{20}\,$eV
decreases with decreasing discrete source density.
Furthermore, magnetic fields $\gtrsim10^{-10}\,$G effectively
block cascade development and more EM energy is channeled into synchrotron
radiation, ending up at GeV-TeV energies. This explains why the photon
to charged cosmic ray ratio also decreases somewhat with increasing EGMF.
Both tendencies are seen by comparing Figs.~\ref{fig4} and~\ref{fig5},
lower panels.

\begin{figure}[ht]
\includegraphics[width=0.5\textwidth,clip=true]{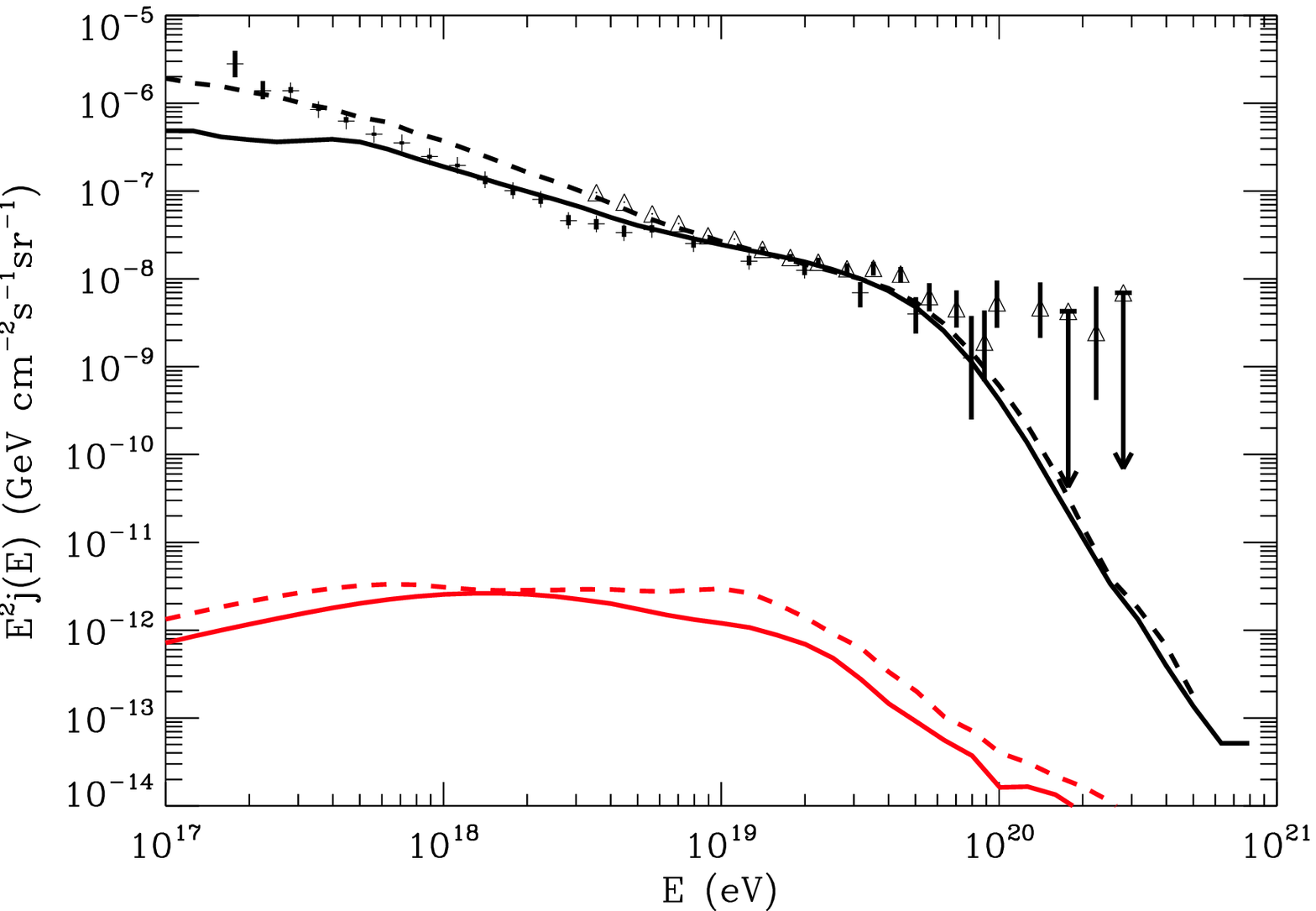}
\includegraphics[width=0.5\textwidth,clip=true]{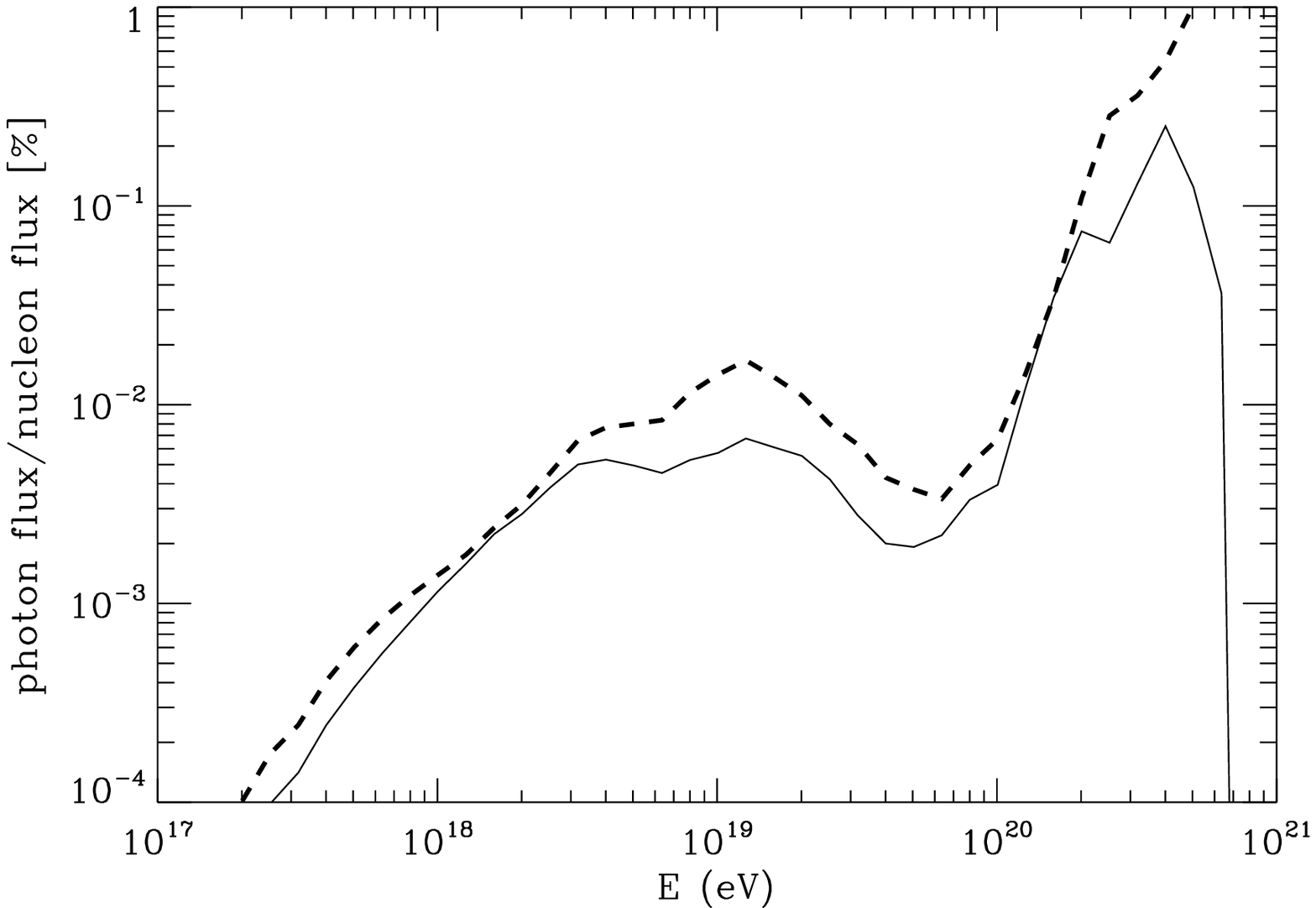}
\caption{Same as Fig.~\ref{fig5} for the discrete source realization,
but now the spectra resulting for negligible EGMF (dashed lines)
and with the EGMF (solid lines) are compared for different injection
parameters, namely $m=0$, $E^{-2.7}$ for the case without EGMF
and $m=3$, $E^{-2.7}$ for the case with EGMF.}
\label{fig6}
\end{figure}

\begin{figure}[ht]
\includegraphics[width=0.5\textwidth,clip=true]{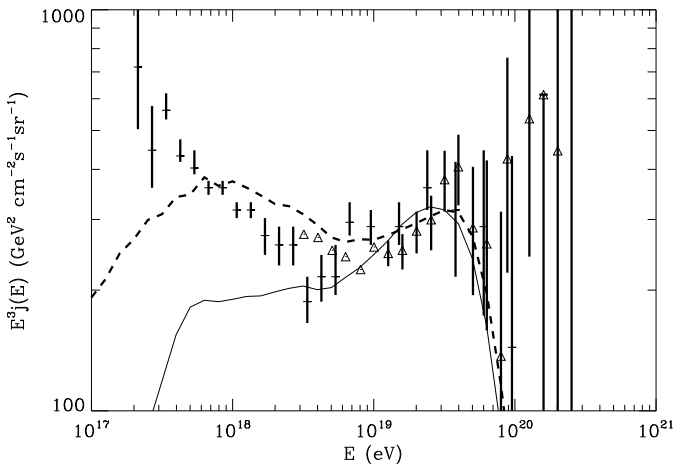}
\caption{Same as Fig.~\ref{fig6}, upper panel, but with the energies
measured by HiRes and AGASA multiplied by a factor of 1.2 and
0.9, respectively, and showing the differential nucleon flux multiplied
by $E^3$.}
\label{fig7}
\end{figure}

In contrast, below $\simeq3\times10^{18}\,$eV, a good fraction of the
photons are produced by pair production of protons at energies
below the GZK threshold. These protons have attenuation lengths
larger than the typical source distance Eq.~(\ref{d_s}).
Furthermore, at these energies the EM attenuation length decreases to
below a few Mpc and the cascade flux thus depends mostly on the
environment of the observer within a few Mpc where the EGMF is negligible,
see Fig.~\ref{fig3}.
These two facts explain why below $\simeq3\times10^{18}\,$eV the
photon to cosmic ray flux ratio is relatively insensitive to discrete
source densities and magnetic environments, as confirmed by
Figs.~\ref{fig4} and~\ref{fig5}, lower panels. In contrast, this
ratio does decrease with increasing magnetic fields around the observer
if these fields are $\gtrsim10^{-10}\,$G, as we will see below for
an EGMF with homogeneous statistical properties.

In Figs.~\ref{fig6} and~\ref{fig7} we compare the spectra in the scenarios with
and without EGMF for different source evolutions, namely $m=3$
and $m=0$, respectively, for an injection spectrum $\propto E^{-2.7}$.
In the case with EGMF, even for strong redshift evolution, a new low-energy,
presumably galactic component becomes necessary below a few $10^{18}\,$eV.
Furthermore, in this case the neutral fractions shown in the lower panel of
Fig.~\ref{fig6} extend considerably below the lower
limit obtained in Ref.~\cite{Gelmini:2005wu,Gelmini:2007sf} which did
not consider structured sources and EGMF.

\begin{figure}[ht]
\includegraphics[width=0.5\textwidth,clip=true]{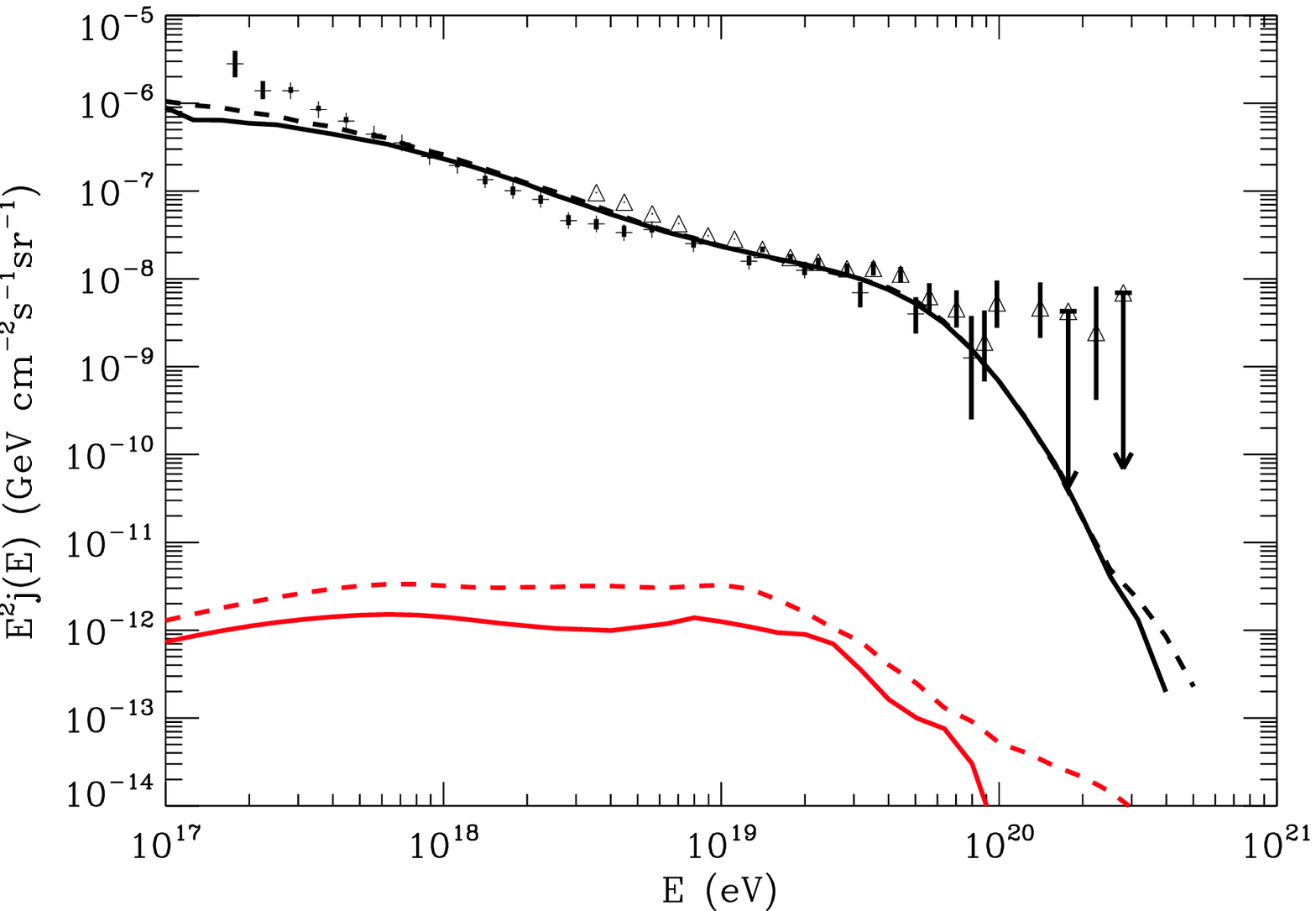}
\includegraphics[width=0.5\textwidth,clip=true]{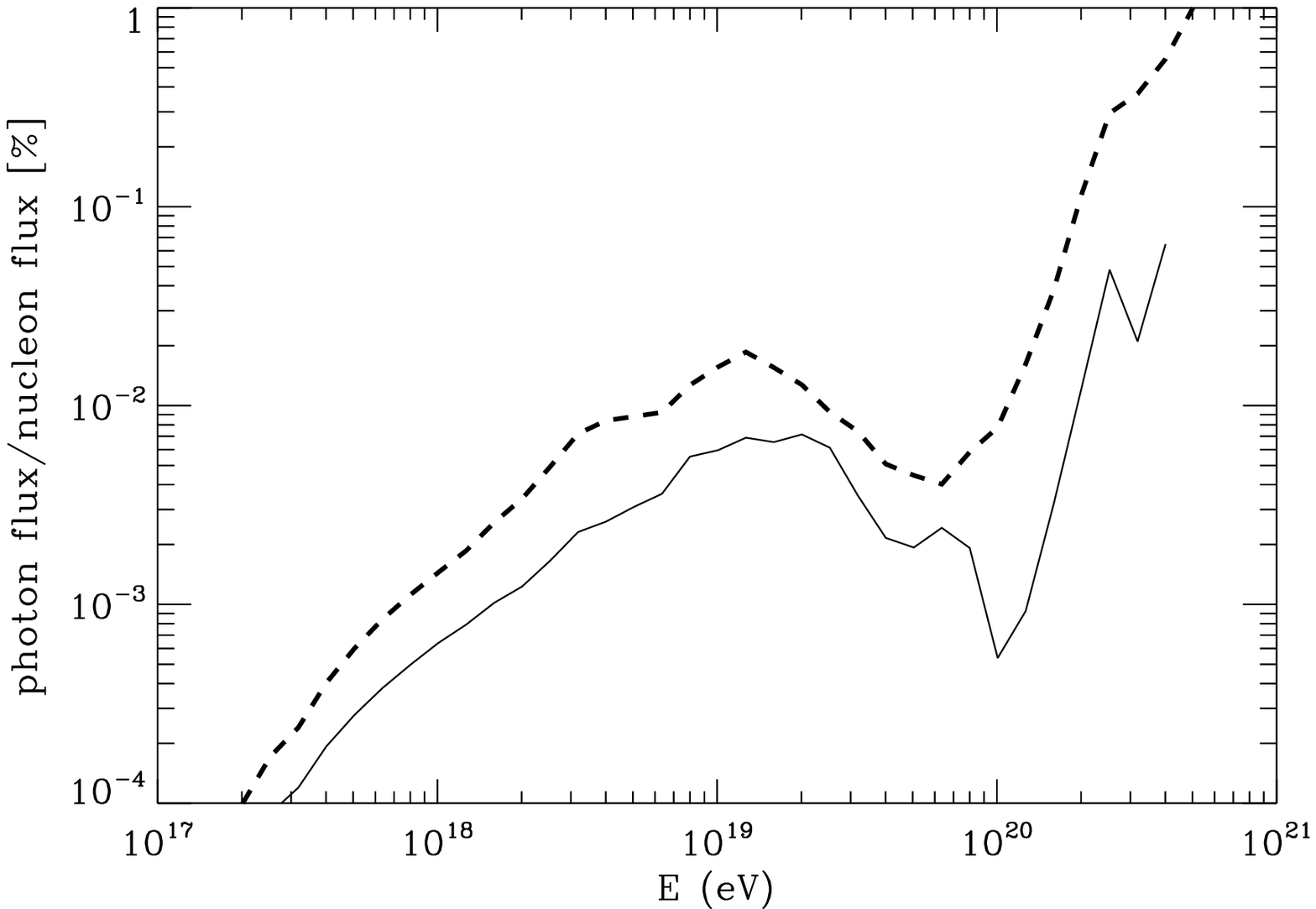}
\caption{As in Figs.~\ref{fig5} and~\ref{fig6} for the discrete
source realization injecting an $E^{-2.6}$ proton spectrum in
the absence of redshift evolution, $m=0$,
but now the spectra resulting for negligible EGMF (dashed lines)
are compared with the spectra for a stochastic EGMF of strength
$10^{-9}\,$G and effective coherence length $\simeq1\,$Mpc
(solid line), see text for details.}
\label{fig8}
\end{figure}

Finally, in order to elucidate the role of the structure of the EGMF,
we consider an EGMF with homogeneous properties: Its Fourier transform
follows a power law spectrum, $|{\bf B}({\rm k})|^2\propto k^{-2.5}$
for $(75\,{\rm Mpc})^{-1}\lesssim k\lesssim(0.6\,{\rm Mpc})^{-1}$, with
an r.m.s. strength of $10^{-9}\,$G. This mimicks the scenarios
considered in Ref.~\cite{Lemoine:2004uw} of an EGMF with a coherence length
$\simeq1\,$Mpc. The field strength is at the high end of realistic
values, but still consistent with current upper limits~\cite{Blasi:1999hu}. 
Fig.~\ref{fig8} compares the resulting spectra with
the case of negligble EGMF. Comparing the upper panels of Figs.~\ref{fig5}
and Fig.~\ref{fig8}, we see that
even in this relatively strong EGMF with homogeneous properties,
the proton flux suppression due to the magnetic horizon effect is
considerably more modest than the confinement effect for the structured
EGMF envisaged here.

In the lower panel of Fig.~\ref{fig8} we see that an EGMF of strength $\gtrsim10^{-10}\,$G
with homogeneous properties tends to suppress the photon to cosmic ray
flux ratio also at energies below $\simeq3\times10^{18}\,$eV, as was
found already in Ref.~\cite{Lee:1995tm} and also seen in Ref.~\cite{Gelmini:2005wu,Gelmini:2007sf}. As remarked
above, this is to be expected because the EM cascade development at
these energies mostly depends on the magnetization of the environment
of the observer within a few Mpc.

Note that the injection spectra needed in the scenarios discussed here where
the cross over to extra-galactic cosmic rays occurs at a few $10^{17}\,$eV
is much steeper than predicted by most acceleration scenarios,
but could be explained as an effective spectrum obtained by averaging
over sources with different maximal energies and harder individual
spectra~\cite{Kachelriess:2005xh}.

The proton injection power required for the scenarios shown in 
Figs.~\ref{fig5}-\ref{fig7} is
$\sim7\times10^{37}\,{\rm erg}\,{\rm s}^{-1}\,{\rm Mpc}^{-3}$
for negligible EGMF and about a factor 10 higher,
$\sim6\times10^{38}\,{\rm erg}\,{\rm s}^{-1}\,{\rm Mpc}^{-3}$,
in the presence of the EGMF considered here. The number for
negligible fields is consistent with the power obtained in
Ref.~\cite{Berezinsky:2007kz} which only considered relatively
weak, unstructured EGMF. The orders of magnitude are
consistent with acceleration in active galactic nuclei, for example.

\begin{figure}[ht]
\includegraphics[width=0.5\textwidth,clip=true]{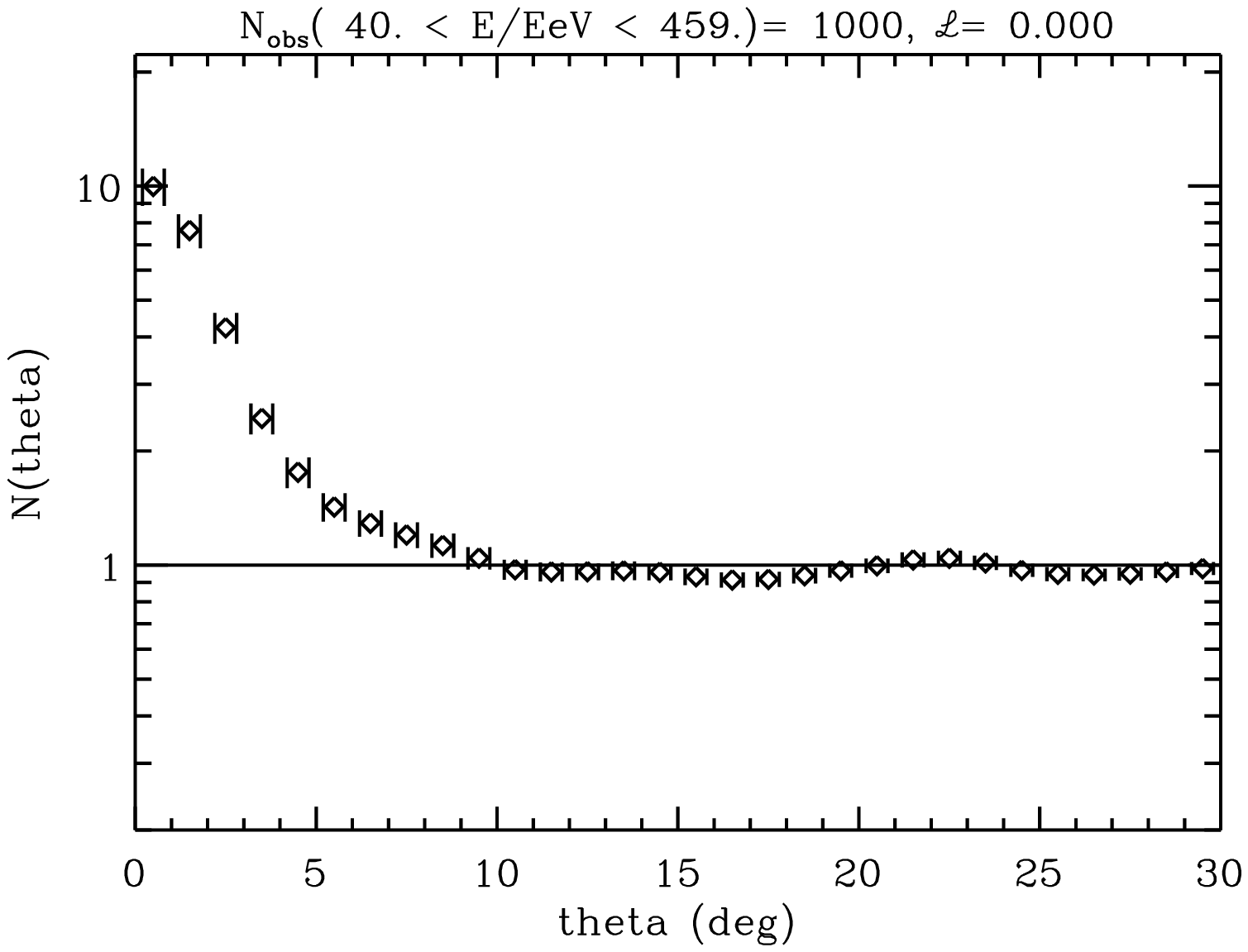}
\includegraphics[width=0.5\textwidth,clip=true]{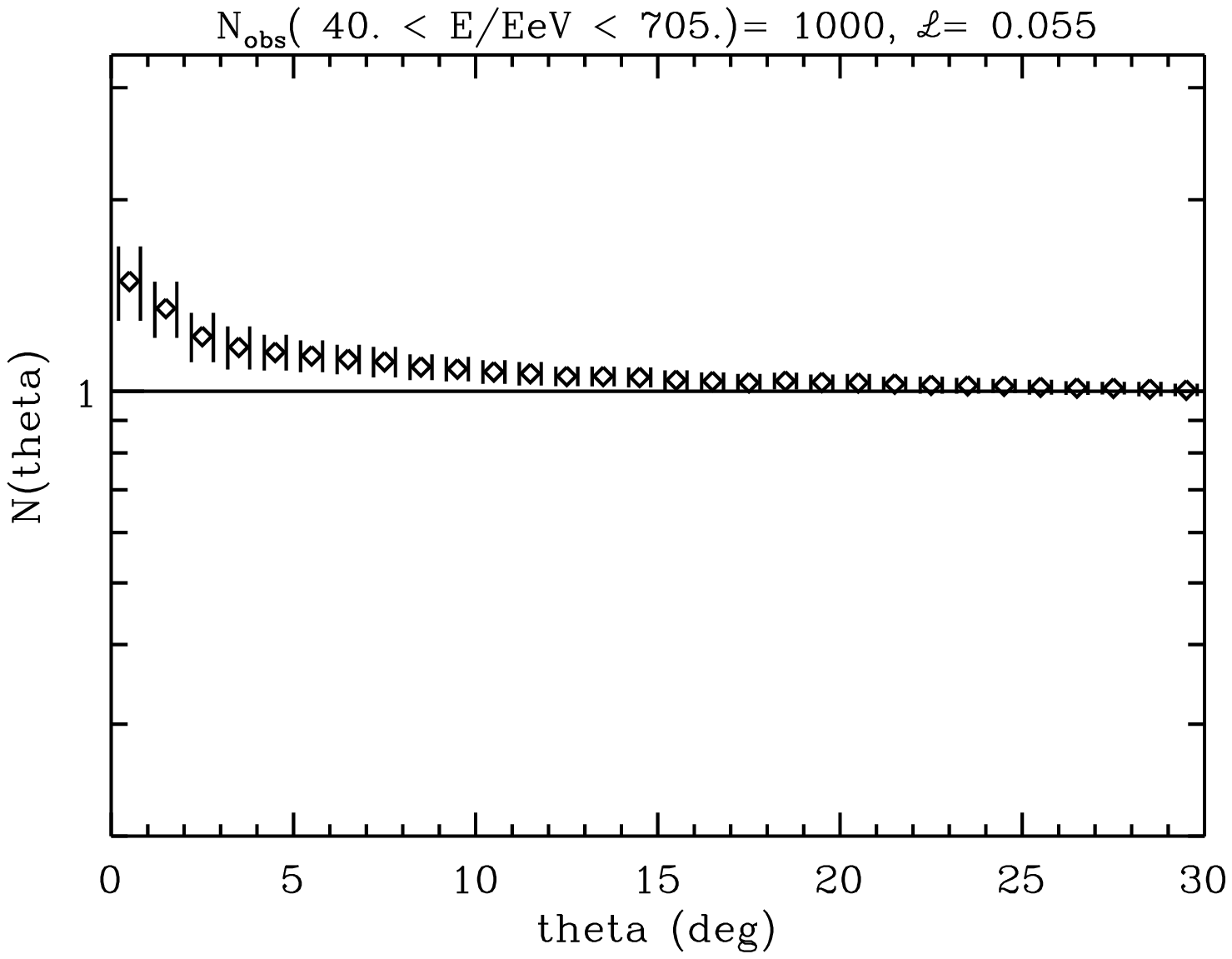}
\caption{Autocorrelation function of UHECR in the discrete source
scenarios shown in Figs.~\ref{fig5} and~\ref{fig6}, assuming $10^3$
events observed above 40 EeV, a factor $5-10$ above the current
world data set. An angular resolution of $1^\circ$ was assumed.
Upper panel: No EGMF. Lower panel: Assuming the EGMF from the LSS
simulation.}
\label{fig9}
\end{figure}

All the scenarios discussed here are consistent with present data
in terms of large scale isotropy and auto-correlations. However,
an increase of the current world exposure by a factor of a few
predicts significant large scale anisotropy as well as small-scale
clustering, at least in the case of modest EGMF. Absence of such
effects would hint to a relatively strong EGMF, as demonstrated in
Fig.~\ref{fig9}, or to a considerably higher source density.

We also found that the diffuse cosmogenic neutrino flux depends mostly
on the source evolution, but little on the source distribution
or EGMF. This is understandable because the neutrino energy attenuation
length is basically the Hubble radius and thus always larger than
all other scales involved. Sources up to cosmological
distances thus contribute to the diffuse neutrino flux. Since neutrinos are
produced by pionproduction of nucleons off the CMB and IR backgrounds,
for a given UHECR injection spectrum, the fraction of the UHECR energy 
transformed to neutrinos should, therefore,
be rather independent of source distribution or EGMF. Our neutrino
fluxes are comparable to other calculations in the literature, for
example Ref.~\cite{DeMarco:2005kt}. A possible enhancement of cosmogenic
neutrino fluxes from the source environment could come from an
enhancement of the IR background around sources such as galaxy clusters,
an effect we have not taken into account in the present work.
One case in which this possibility was considered~\cite{DeMarco:2005va}
does, however, not give much higher diffuse neutrino fluxes when the
constraint from the primary UHECR flux is taken into account.

\section{Uncertainties in the photon to charged cosmic ray flux ratio}
If the actual but poorly known URB is larger than the minimal URB we used
in the simulations presented in the previous section, photon absorption
will be stronger, further reducing the photon to cosmic ray fraction.

A significant fraction of heavier nuclei of atomic number $A$ in the
primary UHECR flux is likely to reduce the photon to charged cosmic ray
flux ratio above $\sim10^{19}\,$eV: Pion-production is shifted down
by a factor $\simeq A$ in energy, falling below $\sim10^{19}\,$eV
already for moderate $A$. An additional source of $\gamma-$rays is
from photo-spallation of nuclei~\cite{Anchordoqui:2006pd}.
But these photons have an energy of $\sim\Gamma_A\times$ a few MeV
which for Lorentz factors $\Gamma_A\lesssim10^{11}$ is below $\sim10^{18}\,$eV.
We recall in this context that the
cosmogenic neutrino flux is also believed to considerably depend
on primary cosmic ray 
composition~\cite{Hooper:2004jc,Anchordoqui:2007tn,Ave:2004uj}.

On the other
hand, additional photons could be produced in nucleon interactions
with an IR background that could be enhanced around sources such as
galaxy clusters, or by nucleon interactions with the ambient baryon
gas in the sources. Both these possibilities have not been taken into
account in the present work. However, we do not expect that these effects
could significantly increase the photon fraction above $\sim10^{15}\,$eV
for the following reasons: First, as discussed in Sect.~III above,
since in this energy range, the EM cascade attenuation length is
smaller than or comparable to the typical source distance, the photon
fraction depends mostly on EM secondary production outside the
sources. Second, the primary cosmic ray flux is not expected to be
significantly modified by these processes, since only about 1\% of
its energy is expected to be converted by interactions with an
enhanced IR background within galaxy clusters~\cite{DeMarco:2005va},
and since the optical depth for $pp$ interactions
is less than unity in galaxy clusters~\cite{Armengaud:2006dd}.

\section{Conclusions}
We have performed simulations following ultra-high energy nucleon
trajectories above $10^{17}\,$eV and compared scenarios with continuous
and discrete source distributions as well as the case of negligible
and highly structured extragalactic magnetic fields reaching microGauss
levels in galaxy clusters. Powerful ultra-high energy cosmic ray sources
likely have densities of order
$10^{-5}\,{\rm Mpc}^{-3}$. We find two considerably non-universal
effects that depend on source distributions and magnetic fields:
We found that the primary extra-galactic cosmic ray flux can become
strongly suppressed below a few $10^{18}\,$eV if the sources are
immersed in highly structured magnetic fields. This is due to the
fact that cosmic ray primaries start to be magnetically confined
and do not reach the observer any more at these energies. This effect
can indeed be considerably stronger than for stochastic magnetic fields
with homogeneous properties and r.m.s. strength $\lesssim10^{-9}\,$G.
We also found that the secondary photon to primary cosmic ray flux ratio
between $\simeq3\times10^{18}\,$eV and $\simeq10^{20}\,$eV is of the
order $10^{-4}$ and
decreases with decreasing source density and increasing magnetization.
In principle at least, this ratio could therefore serve as an
independent measure of these poorly known quantities.
In contrast, the photon to cosmic ray flux ratio is rather
insensitive to source distributions and magnetic
fields at lower energies as well as above $\simeq10^{20}\,$eV where
it can reach the percent level. We also pointed out that a significant
contribution of nuclei heavier than hydrogen to the primary cosmic
ray flux tends to reduce the $\gamma-$ray to charged cosmic ray ratio
further. Also, additional photon production within the sources is
unlikely to increase this ratio significantly.
As a consequence, in acceleration scenarios for the origin
of highest energy cosmic
rays the fraction of secondary photons may be difficult to detect even
for experiments such as Pierre Auger. In contrast, the cosmogenic
neutrino flux
does not significantly depend on source density and magnetization,
but mostly on the redshift evolution of the sources.

\begin{acknowledgments}
This work partly builds on earlier collaborations with Eric Armengaud
and Francesco Miniati. We thank Eric Armengaud and
Markus Risse for stimulating exchanges and useful comments.
During finishing this manuscript, we became aware of a draft by Gelmini,
Kalashev and Semikoz~\cite{Gelmini:2007sf} which studies the $\gamma-$ray to 
cosmic ray ratio in a one-dimensional setting.

\end{acknowledgments}

\end{document}